\documentclass[%
reprint,
superscriptaddress,
 amsmath,amssymb,
 aps,
prb,
]{revtex4-2}

\usepackage{xcolor}
\usepackage{graphicx}
\usepackage{dcolumn}
\usepackage{bm}

\begin{document}

\preprint{APS/123-QED}

\title{Observation of thermally activated coherent magnon-magnon coupling in a magnonic hybrid system}

\author{Dinesh Wagle}
\affiliation{Department of Physics and Astronomy, University of Delaware, Newark, DE 19716, USA}
\author{Yi Li}
\affiliation{Materials Science Division, Argonne National Laboratory, Argonne, Illinois 60439, USA}
\author{Mojtaba Taghipour Kaffash}
\affiliation{Department of Physics and Astronomy, University of Delaware, Newark, DE 19716, USA}
\author{Sergi Lendinez}
\affiliation{Department of Physics and Astronomy, University of Delaware, Newark, DE 19716, USA}
\affiliation{Center for Advanced Microstructures and Devices, Louisiana State University, Baton Rouge, LA 70806, USA}
\author{Mohammad Tomal Hossain}
\affiliation{Department of Physics and Astronomy, University of Delaware, Newark, DE 19716, USA}
\author{Valentine Novosad}
\affiliation{Materials Science Division, Argonne National Laboratory, Argonne, Illinois 60439, USA}
\author{M. Benjamin Jungfleisch}
\email{mbj@udel.edu}
\affiliation{Department of Physics and Astronomy, University of Delaware, Newark, DE 19716, USA}
\date{\today}

\begin{abstract}
We experimentally demonstrate strong magnon-magnon coupling by thermal spin excitations in yttrium iron garnet/permalloy (YIG/Py) hybrid structures using microfocused Brillouin light scattering -- an optical technique that enables the detection of zero-wavevector and higher-order wavevector spin waves in a broad frequency range. The thermally activated magnons in the bilayer lead to a hybrid excitation between magnon modes in the conductive Py layer with a wide wavevector range and the first perpendicular standing magnon modes in the insulating YIG layer, facilitated by strong interfacial exchange coupling. 
To further investigate this coupling, we compare the thermal magnon spectra with the results obtained from electrical excitation and detection methods, which primarily detect the uniform Py mode. 
The realization of coherent coupling between incoherent (thermal) magnons is important for advancing energy-efficient magnonic devices, particularly in classical as well as quantum spin-wave computing technologies.
\end{abstract}

\maketitle


\section{\label{sec:intro} Introduction}
Coupled dynamic hybrid systems can be obtained by combining different physical systems with complementary functionalities and excitations. These dynamic hybrid systems are particularly relevant for studying strong coupling phenomena and for their potential applications in quantum science and engineering. To this end, the coherent coupling of magnons, the collective spin excitations that propagate in magnetically ordered media \cite{stancil09}, with other quasiparticles like photons and phonons has gained increased attention in recent years \cite{avc15,mhd13, kurizki15, chumak15,Shen_PhysRevLett_2022,ZHANG_MatTodayEl_2023, yili20,Awschalom_IEEETrans_2021}. For example, strong magnon-photon coupling enables an effective exchange of energy and information between the magnonic and photonic sub-systems \cite{soykal10, huebl13, xufeng13, tabu14, harder18, Xu_PRL_2024, Wagle_JPM_24}. In these experiments, strong magnon-photon coupling is facilitated by placing a low-damping magnetic material in a high-quality photonic resonator such as a 3D microwave cavity. The extent of the coupling is governed by the coupling strength $ g \propto \sqrt{N} $, where $N$ is total number of spins \cite{dicke54}. Due to the smaller magnetic volume and, hence, the smaller number of spins, the coupling in nanometer-sized media is typically weaker, which poses restrictions on device miniaturization. On the other hand, scaling down of the dimensions is imperative for integrating these devices in real-world applications. An alternative process that could be employed is magnon-magnon coupling, where certain magnon modes couple with other magnon modes (e.g., standing magnon modes). Because magnetic materials are intrinsic magnon resonators, they can be fabricated as thin films while preserving high-quality-factor resonant states, enabling coupling that can be understood as an all-magnonic resonator. As such, the dimensions can drastically be reduced while maintaining strong coupling due to the greater spatial mode overlap~\cite{dai20}.

\begin{figure}[b]
\centering
\includegraphics[width=1\columnwidth]{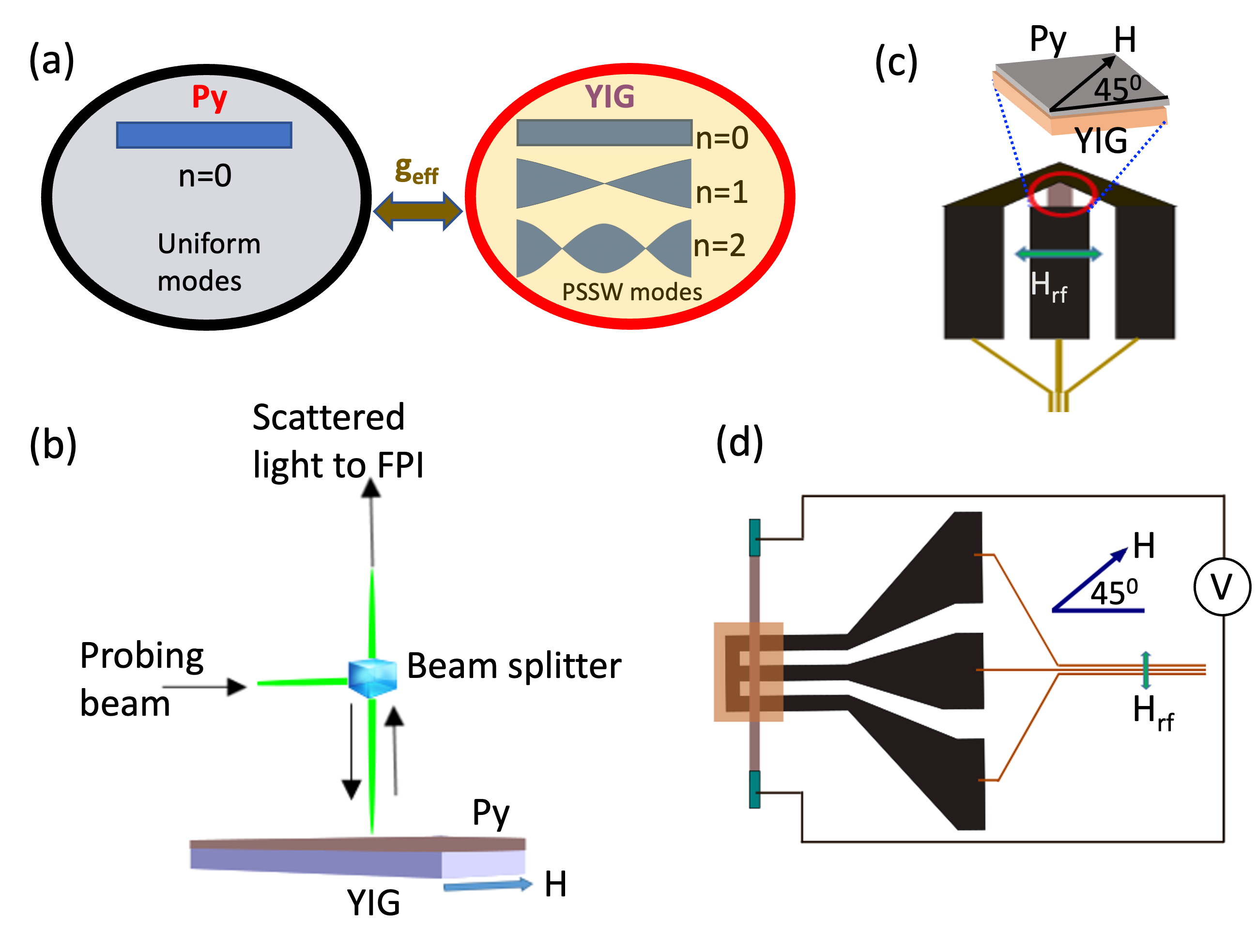}
\caption{(a) Illustration of coupling between uniform Py mode and higher order YIG PSSW modes. (b) Microfocused BLS measurement approach in which the inelastic scattered probe beam is analyzed using a tandem Fabry-P\'{e}rot interferometer (FPI). (c) Spin rectification measurement where an RF current is passed through a conducting Py layer and the rectified signal is measured by a lock-in technique. (d) Spin pumping measurement where RF field is supplied through a coplanar waveguide (shown in black) and the voltage is measured between the two ends of the sample. 
}
\label{setup}
\end{figure}

\begin{figure*}[t]
\centering
\includegraphics[width=1\textwidth]{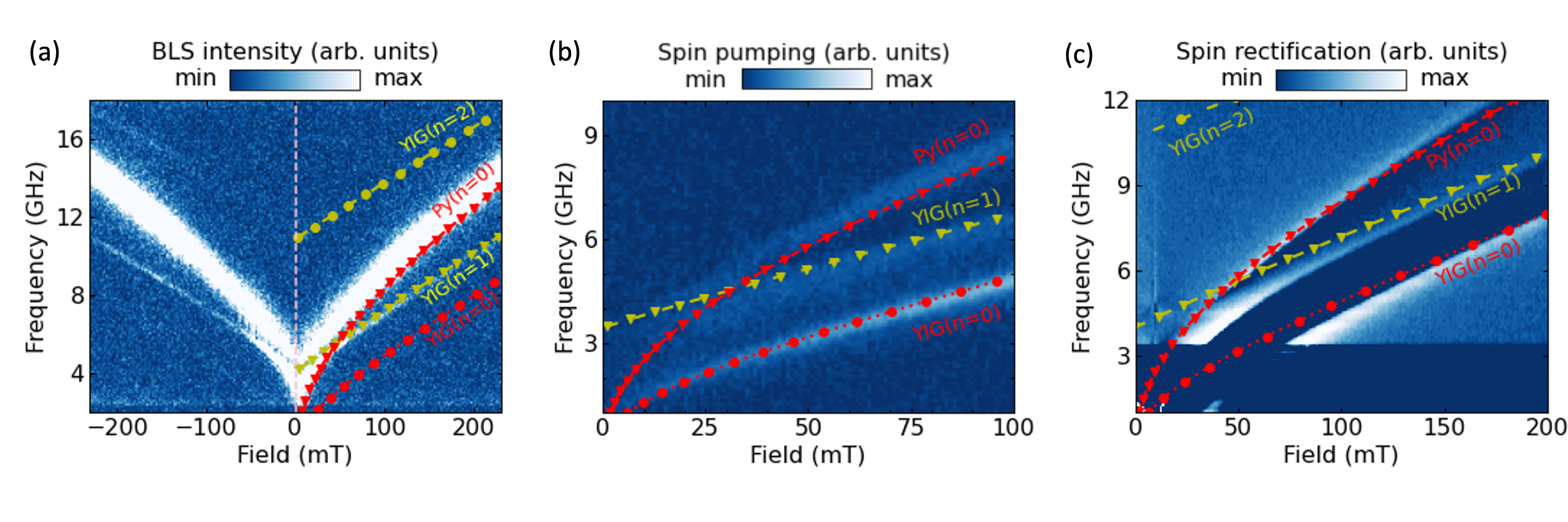}
\caption{Uniform and higher order PSSW modes for YIG and Py obtained using (a) thermal BLS, (b) spin-pumping, and (c) spin rectification techniques for a YIG(84)/Py(10) sample as a function of applied field and frequency. The observed avoided crossing is due to a coherent coupling between the Py mode and YIG ($n=1$) PSSW modes in each of the three figures. The red and yellow curves are fits obtained from Eqs.~(1) and (2).}
\label{fig2}
\end{figure*}

In the past, magnon-magnon coupling experiments have primarily focused on the coherent coupling of magnon modes through interlayer exchange coupling in magnetic thin-film bilayers resonantly excited by a microwave signal \cite{yu18,klinger18,li20,yi21,xiong20,mac19,Makhira_NatCom_2021,Wang_NatCom_2024}. In these studies, strong interlayer magnon-magnon coupling  was observed in an on-chip nano-magnonic device where the coupling between the ferromagnetic resonance (FMR) mode of ferromagnetic nanowires (Ni or Co) and higher-order in-plane standing spin waves in yttrium iron garnet (YIG) thin films was achieved \cite{yu18}. Similarly, such coupling was observed in YIG/Co heterostructures \cite{klinger18, kartik21} and in synthetic antiferromagnets \cite{sud20, chang21, yoi20, Liang_PhysRevB_2025}. Furthermore, the \textit{entanglement} of two magnon modes was studied in two separated YIG spheres coupled to a microwave cavity mode \cite{jie19, agarwal19}. More recent investigations focused on coherent spin pumping arising from the coupled magnetization dynamics of a magnetic thin-film bilayer \cite{li20, yi21}. 
These previous studies investigated coherent magnon-magnon coupling between the uniform mode of one layer with the 
perpendicular standing spin-wave (PSSW) mode of a second layer. PSSWs are usually decoupled from direct microwave excitation due to their mode profile and can therefore only be probed indirectly, for example, in a coupling experiment. For this purpose, the uniform mode of the first layer is coherently excited with an external microwave signal. When the frequency of the uniform mode and the PSSW are degenerate and they are strongly coupled, an avoided level crossing is observed. This approach also enables the indirect detection of PSSWs using microwave excitation. However, these previous studies were limited to a coupling of the uniform, zero-wavevector mode with PSSWs.

Here, we unveil the coherent coupling of incoherently excited -- i.e., thermal -- magnon modes with non-zero wavevector with PSSWs in  a hybrid structure made of nanometer-thick YIG/ permalloy (Ni$_{80}$Fe$_{20}$, Py) bilayers. The hybrid structure is probed by thermal microfocused Brillouin light scattering (BLS) spectroscopy with which we can detect a range of spin-wave wavevectors across a broad frequency range. BLS is a well-established technique for probing magnetization dynamics  with high sensitivity. We show that thermally activated Py modes with  wavevectors ranging form $0-16$~rad/$\mu$m coherently couple to the thermally populated PSSW in  YIG, as is schematically shown in Fig.~\ref{setup}(a). This coupling leads to a characteristic avoided mode crossing indicative for a coherent hybridization of modes. 
The thermal measurements are compared to electrical excitation and detection akin to (1) ferromagnetic resonance driven spin pumping combined with spin rectification in the Py layer and (2) spin-orbit torque ferromagnetic resonance where a microwave current is passed directly through the conductive Py layer and resonance is subsequently detected by spin rectification in the Py layer~\cite{Jungfleisch_Nano_2017}. Control experiments with an insulating SiO$_2$ spacer layer between YIG and Py demonstrate that the observed hybridization arises from interfacial exchange coupling between the layers.

\section{\label{sec:exp} Experimental configuration and setup}
YIG(t$_\mathrm{YIG}$)/Py(t$_\mathrm{Py}$) magnetic thin film bilayer samples of lateral size 44 $\mu$m $\times$ 14 $\mu$m with thickness t$_\mathrm{YIG}$ and t$_\mathrm{Py}$ were grown by magnetron sputtering with a base pressure of 10$^{-8}$ Torr. These samples were used for thermal BLS and for spin rectification measurements. The spin pumping samples were slightly larger, 150 $\mu$m $\times$ 10 $\mu$m. First, the YIG stripe was lithographically defined and grown on a blank Gd$_3$Ga$_5$O$_{12}$ (GGG) substrate, followed by high-temperature annealing in air at 850$^\circ$C for 3 hours after liftoff. Then, the Py layer was defined by another photolithography process. The YIG surface was cleaned by 1-min in-situ ion milling right before the Py growth without breaking the vacuum in order to ensure good interfacial exchange coupling. Last, Ti(5 nm)/Au(100 nm) electrodes were patterned by photolithography and deposited by {electron-beam} evaporation. For every photolithography process, a liftoff resist (LOR) was used to minimize edge roughness of the devices.

We use three complementary experimental techniques: microfocused Brillouin light scattering \cite{Sebastian_BLS} and electrical characterization techniques akin to spin pumping and spin rectification \cite{Harder_PRB_2011}. The three techniques are described in the following.
Figure~\ref{setup}(b) shows the microfocused BLS setup, in which thermal spin dynamics are probed by inelastic scattering of a monochromatic laser light (532 nm) and subsequent frequency analysis using a tandem Fabry-P\'erot interferometer. We note that a high-numerical aperture objective (NA=0.75) was used limiting the detectable spin-wave wavevector to k$^{max}_{sw}$ = 17.8 rad/$\mu$m (spin-wave wavelength of $\lambda^{min}_{sw}$ = 350 nm). 

To complement the thermal measurements capable of detecting a wider wavevector range, we use typical spin pumping and spin-torque ferromagnetic resonance excitation and detection schemes~\cite{Liu_PRL,Sklenar_PRB}, which primarily excite/detect spin dynamics near $k=0$, i.e., the uniform precession. In the spin-torque ferromagnetic resonance approach, we apply a microwave current through the Py layer and use a bias-T to simultaneously detect the  rectified $dc$ signal across ends of the conductive Py layer using a lock-in amplifier, see Fig.~\ref{setup}(c). In the spin-pumping configuration a \textit{nonlocal} detection is used: As is shown in Fig.~\ref{setup}(d), we use a coplanar waveguide (CPW) in the center of the YIG/Py wire to excite dynamics and drive the hybrid system to resonance. The CPW is patterned on the top of the YIG/Py bilayer which is electrically isolated by a thin SiO$_2$ layer. Upon achieving the condition of ferromagnetic resonance, the precessing spins in both ferromagnetic layers mutually pump spin currents in the respective adjacent layers. The signal is rectified in the conducting Py layer and measured by a voltmeter. For both microwave-driven approaches, the biasing magnetic field $H$ is applied at an in-plane angle of $ \theta = 45^\circ$ to maximize the signal strength (Voltage~$\sim$ sin(2$\theta$)) \cite{sklenar15}.

\section{Results and Discussion}

The experimental results obtained for a YIG(84)/Py(10) (numbers in parentheses refer to thickness in nanometers) bilayer are shown in Fig.~\ref{fig2}. Experimental results for YIG(67)/Py(10) and YIG(95)/Py(10) samples are presented in the Supplementary Materials (SM). The thermal BLS spectrum (without any microwave excitation) as a function of applied field is depicted in Fig.~\ref{fig2}(a) in a false-color coded scale. The color represents the BLS intensity, which is directly proportional to the thermal magnon population at a given field. As seen in the figure, four magnon bands -- symmetrically distributed around zero field and corresponding to YIG and Py -- are detected, with maximum intensity shown in white. Their frequencies increase with an increase in the external field and the bands do not cross each other. Instead, a hybridization of modes is observed as indicated by an avoided level crossing between two bands at $\sim$5.7 GHz and a field magnitude of $\sim$50 mT -- a key signature of coherent coupling \cite{li20}.

To gain further insights into the coherent coupling of thermally activated magnons, we compare the thermal microfocused BLS measurements to microwave-driven excitation techniques. The nonlocal spin pumping intensities are shown in Fig.~\ref{fig2}(b), while the spin rectification signals are presented in  Fig.~\ref{fig2}(c) as a function of external bias field and excitation frequency for the YIG(84)/Py(10) bilayer sample. Overall, the results of both measurement techniques agree well with the thermal BLS results. 
{However, we note that the $n=2$ YIG mode is not detected in the microwave experiments. This can be understood by the fact that the net torque on the magnetization for a uniform microwave signal is zero for even modes, while the thermal BLS measurements do not require an external drive.}

\begin{figure}[t]
\centering
\includegraphics[width=0.99\columnwidth]{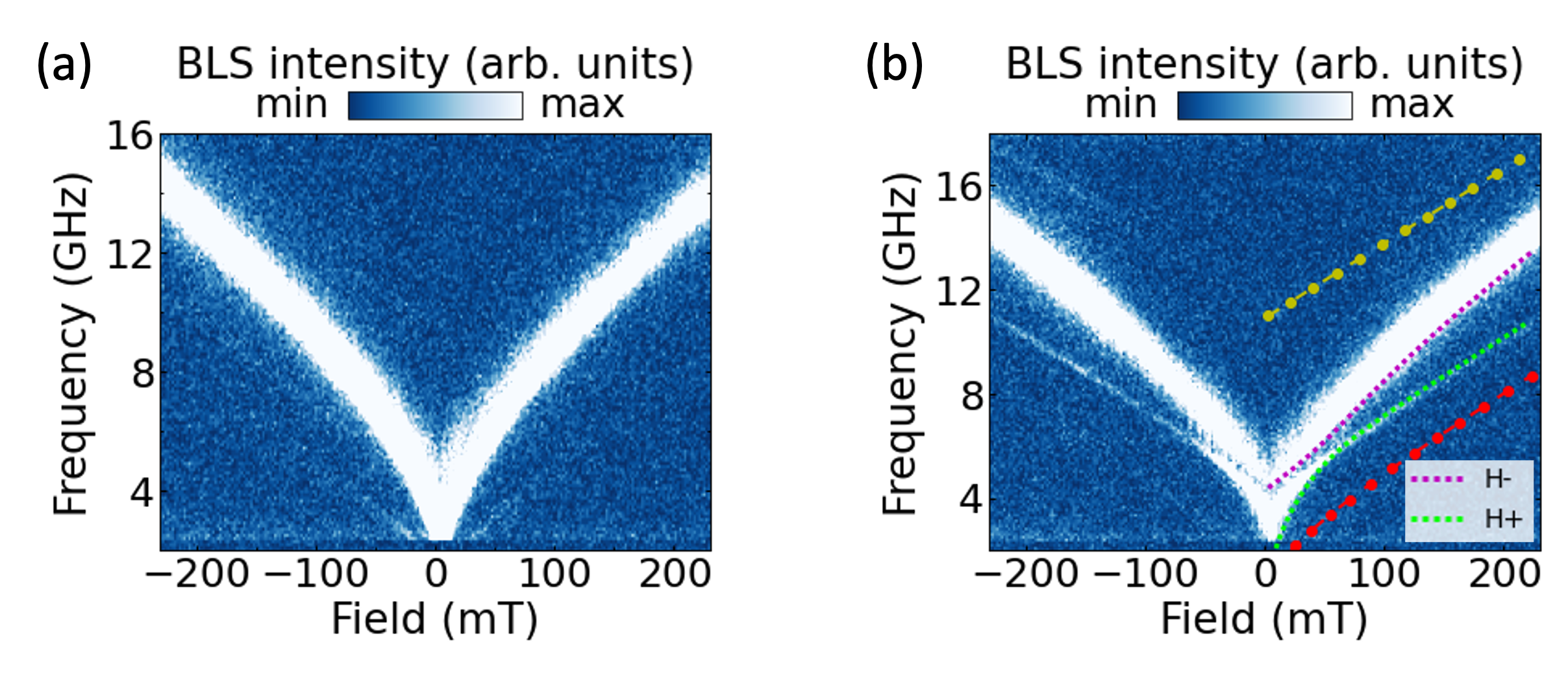}
\caption{(a) BLS spectrum for the YIG(86)/SiO$_2$(5)/Py(10) sample, in which SiO$_2$ disables the direct exchange interaction between YIG and Py. As a result, we do not observe any mode coupling. (b) Corresponding BLS data for YIG(84)/Py(10) without the SiO$_2$ interlayer, the same data as presented in Fig.~\ref{fig2}(a), with the observed avoided crossing between the two coupled modes modeled by two coupled harmonic oscillator curves, which are overlaid as colored dashed lines and labeled as H$+$ and H$-$. Curves in red and yellow correspond to YIG ($n = 0$) and YIG ($n = 2$) modes, respectively. 
}
\label{hybrid}
\end{figure}

We also performed the thermal BLS measurements on a YIG(86)/SiO$_2(5)$/Py(10) trilayer, where a sputtered SiO$_2$ spacer was inserted to prevent direct contact between the YIG and Py layers, and hence, turn off interfacial exchange interaction. The corresponding spectra are presented in Fig.~\ref{hybrid}(a). If the exchange coupling between YIG and Py is turned off by inserting SiO$_2$ between the layers, no avoided crossing is observed. This contrasts the behavior observed for the YIG/Py system where YIG and Py are in direct contact [see Fig.~\ref{fig2}(a) or \ref{hybrid}(b)]. 

The magnetic resonance of the YIG and Py thin films for the uniform modes can be modeled using the Kittel equation \cite{kittel48}:
\begin{equation}
    \mu_0H_r = \frac{-\mu_0M_s+\sqrt{(\mu_0M_s)^2+4\frac{\omega^2}{\gamma^2}}}{2},
\label{kittel}    
\end{equation}
where H$_r$ is the resonance field, M$_s$ is the saturation magnetization, $\omega$ is the FMR frequency and $\gamma$ is the gyromagnetic ratio.\\
For higher order spatially non-uniform PSSW modes, an effective field, $\mu_0H_{ex}(k) = (2A_{ex}/M_s)k^2$ lowers the resonance field H$_r$, and Eq.~(\ref{kittel}) in case of 
PSSW becomes \cite{kittel51, kali86}:
\begin{equation}
    \mu_0H_r = \frac{-\mu_0M_s+\sqrt{(\mu_0M_s)^2+4\frac{\omega^2}{\gamma^2}}}{2} - \frac{2A_{ex}}{M_s}k^2,
\label{pssw}    
\end{equation}
where A$_{ex}$ is the exchange stiffness, $k$ = $n\pi/t$; $n$ labels the index of PSSW modes [Fig.~\ref{setup}(a)], and $t$ is the film thickness.


Using Eq.(\ref{pssw}), we model the experimental results by fitting  the data for $n$ = 0, 1, 2. This fitting yields the following parameters: $\mu_0$M$_s$(YIG) = 210 mT, $\mu_0$M$_s$(Py) = 810 mT and A$_{ex}$ = 4.4~pJ/m. The values are in agreement with the literature~\cite{li20}. The obtained curves are plotted on top of the experimental color-coded data in Fig.~\ref{fig2}. The three observed modes in all three figures correspond to the uniform mode (n=0) and the first-order PSSW mode ($n=1$) for YIG, along with the uniform mode ($n=0$) for Py. Additionally, in Fig.~\ref{fig2}(a), a fourth mode corresponding to the second-order PSSW mode ($n=2$) for YIG is observed although the signal strength is drastically reduced.

{In each measurement, we identify an avoided crossing between the YIG (n=1) mode and the uniform Py mode at $\sim$5.7 GHz and a field magnitude of $\sim$50 mT, precisely where they would otherwise intersect. This avoided crossing is less apparent in spin rectification measurements due to the more complex lineshape, while it is more distinct in the BLS measurements. This distinction arises because BLS captures contributions from different wavevectors and employs a microscope for detection. The Py mode appears more intense, as the light is directly focused on the Py layer, and its broader linewidth results from thermally activated range of a higher order spin waves, as will be discussed below. Furthermore, the BLS intensity of the YIG ($n=1$) PSSW mode is higher than that of both the uniform YIG mode and the YIG ($n=2$) PSSW mode (see SM Fig. 4). This indicates that the YIG mode coupled to Py is the most prominent, demonstrating that the coupling enhances the detection efficiency of the YIG signal.}

The coupling between the modes can be modeled using two coupled harmonic oscillators, as described by \cite{li20, huebl13, bai15}:
\begin{equation}
    \mu_0H_{\pm} = \mu_0\frac{H_r^{YIG} + H_r^{Py}}{2} \pm \sqrt{\Big(\mu_0\frac{H_r^{YIG} - H_r^{Py}}{2}\Big)^2+g_c^2},
\label{harmonic}    
\end{equation}
where H$_r^{YIG}$ and H$_r^{Py}$ are the resonance fields of YIG and Py, respectively, and g$_c$ represents the interfacial exchange coupling strength. These equations accurately describe the avoided crossing observed between the YIG($n=1$) and uniform Py modes as shown in Fig.~\ref{hybrid}(b). From the fitting, we extract an interfacial exchange coupling strength g$_c$ = 10 mT which is close to the value reported in Ref.~\cite{li20} for coherent magnons. This result demonstrates that magnon-magnon coupling does not require a coherent magnon population or an externally applied coherent drive. Instead, thermally excited magnons alone are sufficient to induce and sustain strong coupling between magnon modes, highlighting the ability of incoherent excitations to mediate hybridization in magnonic systems.
 
\begin{figure}
\centering
\includegraphics[width=0.9\columnwidth]{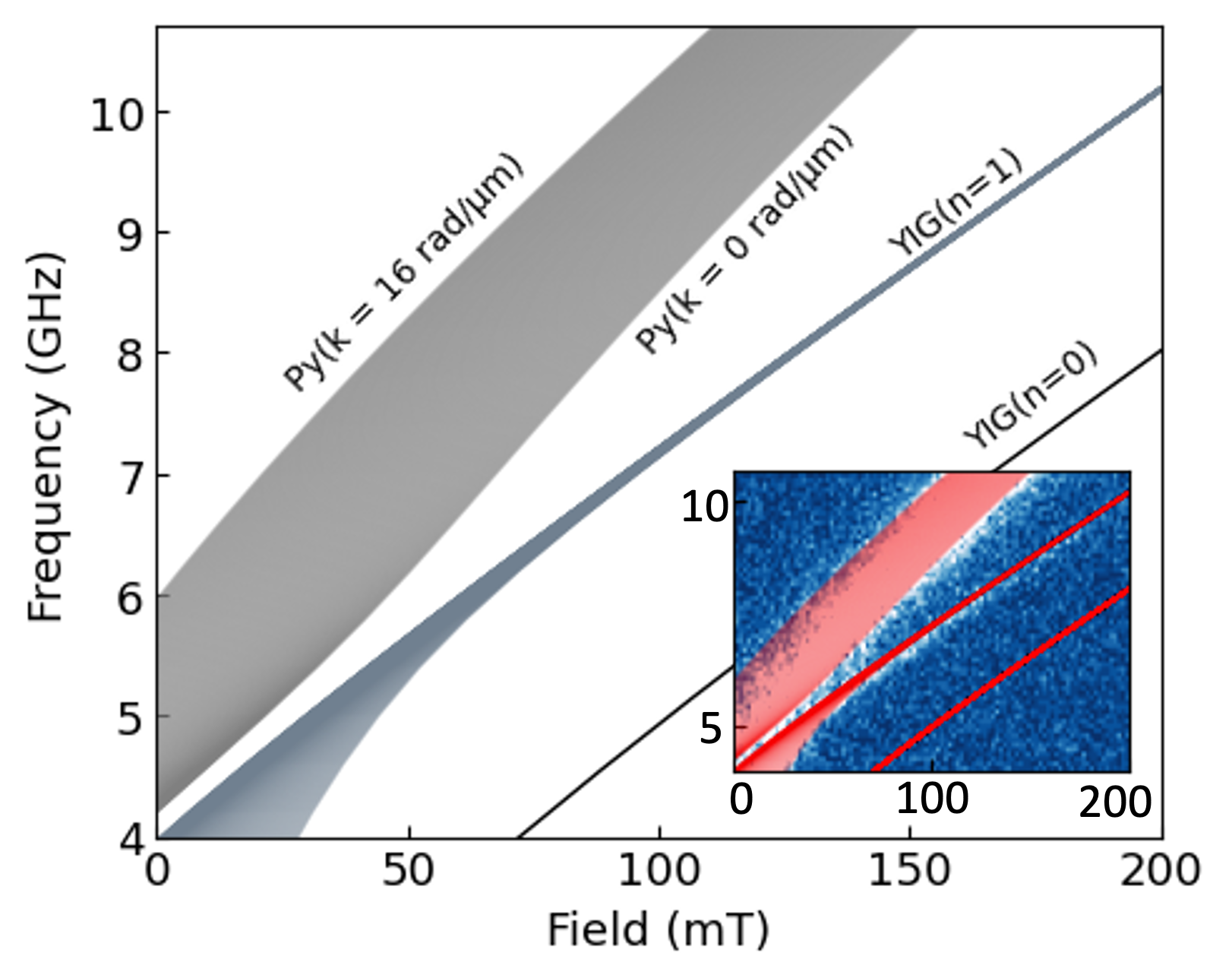}
\caption{The calculated frequency-field relationship illustrating the hybridization between the first order PSSW mode of YIG and the Py spin-wave band spanning a wavevector range from $k$ = 0 to 16 rad/$\mu$m. The inset shows the calculated frequency vs. field curve overlaid on the experimental BLS spectra, highlighting the agreement between theoretical model and experimental observations. 
}
\label{model}
\end{figure}

{
To account for the broad Py band observed in thermal BLS measurements, we model the coupling by considering higher wavevectors for the Py layer. The highest-frequency mode (at a given field) observed in Fig.~\ref{fig2}(a) corresponds to k $\approx$ 16 rad/$\mu$m magnetostatic surface spin wave (MSSW) or Damon–Eshbach waves, as we can see by considering the following equation~\cite{kali86, Kaffash_QST_2022}:
\begin{equation}\label{eq:MSSW}
    f_{MSSW} = \frac{\gamma}{2\pi}\mu_0\sqrt{H_r\Big(H_r+ M_s\Big)+\frac{M_s^2}{4}\Big(1-e^{-2kt}\Big)},
\end{equation}
where $t$ (=10 nm) is thickness of the Py film. 
We model the coupling between the $k = 0 - 16$ rad/$\mu$m MSSW Py band with YIG ($n=1$) PSSW mode using the coupled harmonic equation [Eq.~(\ref{harmonic})]. For that, we assume the same coupling strength $g_c$ = 10 mT as previously determined for the coupling between the YIG ($n=1$) PSSW mode and the uniform Py mode. The results of this model are illustrated in Fig.~\ref{model}. Our model closely reproduces the experimental results, as shown in the inset of Fig.~\ref{model}. The maximum wavevector $k$ $\approx$ 16 rad/$\mu$m corresponds to a spin-wave wavelength of $\lambda$ = 392 nm. This wavevector range matches the theoretical detection limit of our microfocused BLS system (k$^{max}_{sw}$ = 17.8 rad/$\mu$m). We note that the contribution of backward volume spin waves to the broadening of the Py band is negligible, as they have a smaller frequency spread (see SM).}

\section{Outlook}
In conclusion, we showed the coherent coupling of incoherent magnons using thermal BLS measurements and confirmed the experimental results by a theoretical model taking into account magnon-magnon coupling. We found that, due to strong interfacial exchange, the thermal excitation of magnons in the YIG/Py bilayer leads to a coupling between the a range of spin-wave wavectors in the conductive Py layer and higher order PSSW modes in the insulating YIG layer. The thermal BLS spectra agree well with the results obtained by microwave techniques. A broad band of Py magnons with  wavevectors ranging form $0-16$~rad/$\mu$m was observed by BLS, which was absent in the microwave spectroscopy measurements as those techniques primarily couple to the uniform excitation. 
Our results pave the way for magnonic hybrid systems utilizing the coherent coupling of thermal, non-zero wavevector magnons, with potential applications in miniaturized coherent information processing devices and applications \cite{anjan21, yili20}.

\section{Acknowledgement}
We thank Dr. Shulei Zhang, Xavier Moskala, and Dr. Pengtao Shen for helpful discussions. We acknowledge support by the U.S. Department of Energy, Office of Basic Energy Sciences, Division of Materials Sciences and Engineering under Award DE-SC0020308. The authors acknowledge the use of facilities and instrumentation supported by NSF through the University of Delaware Materials Research Science and Engineering Center, DMR-2011824. The fabrication of the YIG/Py devices by Y. L. and V. N. was supported by the U.S. DOE, Office of Science, Basic Energy Sciences, Materials Sciences and Engineering Division under Award DE-SC0022060.

\bibliography{apssamp}

\end{document}